# Giant tunability of magnetoelasticity in Fe$_4$N system: Platform for unveiling correlation between magnetostriction and magnetic damping

Keita Ito[1,*], Ivan Kurniawan[2], Yusuke Shimada[1,†], Yoshio Miura[2], Yasushi Endo[3,4], & Takeshi Seki[1,4]

[1] Institute for Materials Research, Tohoku University, Sendai 980-8577, Japan

[2] Research Center for Magnetic and Spintronic Materials, National Institute for Materials Science, Tsukuba, Ibaraki 305-0047, Japan

[3] Department of Electrical Engineering, Graduate School of Engineering, Tohoku University, Sendai 980-8579, Japan

[4] Center for Science and Innovation in Spintronics, Tohoku University, Sendai 980-8577, Japan

[*] Corresponding author. Email: keita.ito.e3@tohoku.ac.jp

[†] Present address: Department of Advanced Materials Science and Engineering, Faculty of Engineering Sciences, Kyushu University, Fukuoka 816-8580, Japan

Flexible spintronics has opened new avenue to promising devices and applications in the field of wearable electronics. Particularly, miniaturized strain sensors exploiting the spintronic function have attracted considerable attention, in which the magnetoelasticity linking magnetism and lattice distortion is a vital property for high-sensitive detection of strain. This paper reports the demonstration that the magnetoelastic properties of $Fe_4N$ can be significantly varied by partially replacing Fe with Co or Mn. The high quality $Fe_4N$ film exhibits large negative magnetostriction along the [100] direction ($\lambda_{100}$) of −121 ppm while $Fe_{3.2}Co_{0.8}N$ shows $\lambda_{100}$ of +46 ppm. This wide-range tunability of $\lambda_{100}$ from −121 to +46 across 0 allows us to thoroughly examine the correlation between the magnetoelasticity and other magnetic properties. The strong correlation between $\lambda_{100}$ and magnetic damping ($\alpha$) is found. The enhanced extrinsic term of $\alpha$ is attributable to the large two magnon scattering coming from the large magnetostriction. In addition, the density of states at the Fermi level plays a primal role to determine both $\lambda_{100}$ and the intrinsic term of $\alpha$. Thanks to the giant tunability and the bipolarity of magnetoelasticity, magnetic nitrides are candidate materials for high-sensitive spintronic strain sensors.

Flexible electronic devices have rapidly become important in the coming trillion-sensor era, in which wearable devices will spill over. Flexibility is required not only for semiconductor-based electronic devices, but also for spintronic devices, and the vast research field of flexible electronics involves a variety of research topics coming from technological demands and scientific interests, *e.g.* device fabrication and operation on flexible substrates, and improved performance and new functionalities under the application of mechanical strain[1-7]. As we look at the current flexible spintronics, miniaturized strain sensors exploiting the spintronic function[6] have attracted much attention, where the magnetoelasticity linking magnetism and lattice distortion is a vital property for high-sensitive and multi-directional strain detection. Through the inverse magnetostriction effect, the magnetization vector of a magnetic material changes depending on the magnitude and the direction of mechanical stress applied to the magnet. Thus, large magnetoelastic materials and the elucidation of underlying physical mechanism are keys for the development of high-sensitive devices.

Although there are several magnetoelastic materials, the spintronic applications need to show large magnetoresistance (MR) effect as well as large magnetostriction. Tb-Dy-Fe alloys[8] and Fe-Ga alloys[9] are well known ferromagnets exhibiting large magnetostriction. However, since no large MR effect has been reported in any giant MR (GMR) or tunneling MR (TMR) devices with Tb-Dy-Fe or Fe-Ga, these materials are excluded from consideration. The recent first-principles calculation[10] has predicted that $Fe_4N$, a ferromagnetic iron nitride exhibits the magnetostriction constant several times larger than that of typical ferromagnetic metals such as Fe, Co, and Ni, and there is a report on experimental investigation for magnetostriction of $Fe_4N$ although the reported value was smaller than the theoretical prediction[11]. In addition, a large TMR effect in a magnetic tunnel junction[12], anomalous Hall effect[13], and anomalous Nernst effect[13,14] have been reported for systems with $Fe_4N$. These features make ferromagnetic nitrides a promising material group not only for spintronic devices[15,16], but also for flexible spintronic devices. Moreover, epitaxial growth of ferromagnetic nitrides on flexible mica substrates and modulation of magnetic properties by stress application have been reported[17], and the $Fe_4N$ does not contain minor metal elements and rare earth elements, which is another advantage from the viewpoint of large-scale fabrication of wearable devices. Thus, it should be clarified to what extend the magnetoelasticity can be enhanced and controlled for ferromagnetic nitrides in order to achieve best performance of flexible spintronic devices.

Here we report that the magnetoelastic properties of $Fe_4N$ can be significantly varied by partially replacing Fe with Co or Mn. As shown in Fig. 1a, the Co substitution in $Fe_4N$ corresponds to the electron doping whereas the Mn substitution leads to the electron depletion. This variation in the number of valence electrons gives rise to the shift of the Fermi level ($E_F$) position, and also makes the $Fe_4N$ a suitable platform to examine the correlation between the magnetoelasticity and other magnetic properties experimentally and theoretically. By measuring various magnetic properties such as magnetostriction, saturation magnetization ($M_s$), magnetic anisotropy constant ($K_1$), and magnetic damping constant ($\alpha$), we experimentally and theoretically reveal the strong correlation between magnetostriction and magnetic damping.

**Co or Mn substitution in Fe$_4$N films**

Fe$_4$N, Fe$_{4-x}$Mn$_x$N ($x$ = 0.1, 0.4, and 1.0), and Fe$_{4-y}$Co$_y$N ($y$ = 0.2, 0.5, 0.8, 1.7, and 2.2) films with a thickness of approximately 23 nm were grown on SrTiO$_3$(STO)(001) substrates. The exact values of thicknesses, which were estimated from the x-ray reflectivity, were given in the Supplementary Information. Figures 1b, 1c, and 1d display the x-ray diffraction (XRD) patterns for the representative samples of the Fe$_{3.2}$Co$_{0.8}$N, Fe$_4$N, and Fe$_3$MnN films, respectively. The red (blue) lines denote the out-of-plane (in-plane) XRD patterns. The 001 and 002 peaks of nitrides appear in the out-of-plane XRD patterns while the 100 and 200 peaks are observed in the in-plane XRD patterns. These facts indicate the epitaxial growth of the nitride films on the STO(001) substrates. From the structural characterization using XRD as well as reflection high-energy electron diffraction patterns (not shown here), it was confirmed that all the samples are epitaxially grown and no secondary phase exists.

Figure 2a shows the magnetization curves of the Fe$_4$N film measured at room temperature. When an external magnetic field ($H$) is applied in the in-plane [100] direction (solid curve), the remanent magnetization ($M_r$) is equal to $M_s$. On the other hand, $M_r$ is decreased when $H$ is applied in the in-plane [110] direction (dashed curve). This difference originates from the cubic symmetry of magnetocrystalline anisotropy in Fe$_4$N with the [100] easy magnetization axis[18,19]. From the magnetization curves, $M_s$ and $K_1$ were evaluated for all the samples, where $K_1$ was estimated from the area enclosed between the [100] and [110] magnetization curves.

**Magnetoelasticity and magnetization dynamics for nitride films**

Figure 2b illustrates the experimental setup for measuring the magnetostriction: optical cantilever method[20]. A rectangular-shaped sample with the size of 10 mm × 20 mm was fixed on one side and illuminated with a laser beam from the vertical direction of the film surface. A rotational $H$ of up to 175 Oe was applied in the in-plane direction to induce the magnetostrictive effect, and the degree of deflection of the sample was measured by the position of the reflected laser light ($s$). Since $\phi$ is the angle between the [100] direction of the nitrides and $H$, i.e. $\phi = 0°$ is the [100] of nitrides, this measurement setup corresponds to the evaluation of magnetostriction along [100] ($\lambda_{100}$). The $\phi$ dependence of $s$ at $H = 175$ Oe is plotted in Fig. 2c. The obtained sinusoidal curve was numerically fitted and the displacement of $s$ ($\Delta s = s(\phi = 0°) - s(\phi = 90°)$) was obtained, allowing us to evaluate $\lambda_{100}$ with high accuracy (see Supplementary equation 1). Figure 2d shows the $H$ dependence of $\lambda_{100}$ for the Fe$_{3.2}$Co$_{0.8}$N, Fe$_4$N, and Fe$_3$MnN films. At the saturated state of magnetization by applying $H = 175$ Oe, a large magnetostriction, $\lambda_{100} = -121$ ppm, is obtained for the Fe$_4$N film, which is three times larger than the previous report (−40 ppm[11]) in absolute value and is close to the theoretical value of −130 ppm by the first-principles calculation[10]. Also, it is noted that $\lambda_{100} = -121$ ppm is much larger than the values for Fe-Co alloys without high temperature annealing[21], which are representative ferromagnets for the conventional spintronics, and is even in the same order of magnitude as that for the representative magnetostrictive material Fe-Ga. In contrast to Fe$_4$N, Fe$_{3.2}$Co$_{0.8}$N and Fe$_3$MnN exhibit small positive and negative $\lambda_{100}$, respectively.

The magnetization dynamics of the Fe$_4$N, Fe$_{3.2}$Co$_{0.8}$N, and Fe$_3$MnN films is given in Figs. 2e-2h. Figures 2e and 2f plot the $H$ dependence of the ferromagnetic resonance (FMR) frequency ($f_0$) and resonance linewidth ($\Delta f_0$) at the in-plane $H$ applied along the [100] direction, which is called the IP configuration. The inset of Fig. 2e displays the FMR spectrum of the Fe$_4$N at $H = 1400$ Oe. The $H$ dependences of $f_0$ and $\Delta f_0$ are fitted by the following equations:

$$f_0(H) = (\gamma/2\pi)\sqrt{(H + H_{a,ip})(H + 4\pi M_{eff})} \quad \text{and} \quad \Delta f_0(H) = \alpha_{ip}(\gamma/2\pi)(2H + H_{a,ip} + 4\pi M_{eff}) + (\gamma/2\pi)H_{inh},$$

where $\gamma$ is the gyromagnetic ratio, $H_{a,ip}$ is the in-plane magnetic anisotropy field, $M_{eff}$ is the effective magnetization, $H_{inh}$ is the inhomogeneous field, and $\alpha_{ip}$ is the damping parameter estimated in the IP configuration. The fitting of $f_0$ as a function of $H$ was done with the fixed $\gamma$ value of $1.76 \times 10^7$ rad$^{-1}$ Oe$^{-1}$ with the assumption of $g$-factor of 2, which enabled to evaluate $H_{a,ip}$ and $M_{eff}$ as fitting parameters. Those values were used for the fitting of $\Delta f_0$ as a function of $H$. Finally, $\alpha_{ip}$ was evaluated from these IP configuration measurements. Similar analyses were carried out for the results at $H$ applied in the out-of-plane direction (OOP configuration) as shown in Figs. 2g and 2h, where the following equations were used: $f_0(H) = (\gamma/2\pi)(H - H_s)$ and $\Delta f_0(H) = \alpha_{\text{oop}}(\gamma/2\pi)(2H - 2H_s) + (\gamma/2\pi)H_{\text{inh}}$, where $H_s$ is the saturation field of magnetization by the out-of-plane $H$. $\alpha_{\text{oop}}$ corresponds to the intrinsic term of magnetic damping, which is defined as intrinsic $\alpha$ while $\alpha_{\text{ip}}$ involves both intrinsic and extrinsic terms[22,23]. This paper defines $\Delta\alpha = \alpha_{\text{ip}} - \alpha_{\text{oop}}$, and $\Delta\alpha$ expresses the extrinsic contribution of magnetic damping constant. Although one may think the nonlocal spin relaxation process[24] due to the effect of spin pumping to adjacent layers as a component of $\Delta\alpha$, that contribution is negligibly small because of small spin-orbit interaction of Ti capping layer in the present layer stacking.

**Composition dependences of magnetic properties**

Figures 3a, 3b, and 3c plot the $x$ and $y$ dependences of $M_s$, $K_1$, and $\lambda_{100}$, respectively. $M_s$ decreases by the Mn substitution into Fe$_4$N because Fe$_{4-x}$Mn$_x$N exhibits ferrimagnetic spin structures[25]. Although Fe$_{4-y}$Co$_y$N is a ferromagnet[26,27], the Co substitution also reduces $M_s$, which results from the Co atom having a smaller spin magnetic moment than Fe. $K_1$ decreases by the Mn substitution for $x > 0.4$. In the case of Co substitution, $K_1$ decreases with increasing $y$ and its sign changes to negative at $y > 0.75$. The most striking features are found in the composition dependence of $\lambda_{100}$. First, as mentioned above, the large $\lambda_{100}$ of −121 ppm is achieved for the present high-quality Fe$_4$N epitaxial film. Next, the value of $\lambda_{100}$ takes the negative sign for $0 \leq x \leq 1.0$ and $0 \leq y \leq 0.5$, and becomes positive at $y \geq 0.8$. More specifically,

$\lambda_{100}$ values can be modulated over a wide range from −121 ppm for Fe$_4$N to +46 ppm for Fe$_{2.3}$Co$_{1.7}$N with $\lambda_{100}$ close to 0 near $y = 0.75$. This is clear demonstration that the Co substitution in Fe$_4$N realizes the large tunability and bipolarity of magnetoelastic property. Another finding is that there is no strong correlation between $K_1$ and $\lambda_{100}$ (see the Co-rich composition) although both are parameters related to the spin-orbit interaction.

Figures 3d and 3e plot the $x$ and $y$ dependences of $\alpha_{\text{ip}}$ and $\alpha_{\text{oop}}$, respectively. The value of $\alpha_{\text{ip}}$, the sum of intrinsic and extrinsic $\alpha$, shows the maximum value of 0.04 for the Fe$_4$N film and rapidly decreases with the Mn or Co substitution. The value of $\alpha_{\text{oop}}$, corresponding to intrinsic term of $\alpha$, exhibits the composition dependence similar to that of $\alpha_{\text{ip}}$ except for $x = 1.0$ and $y = 2.2$. From the values of $\alpha_{\text{ip}}$ and $\alpha_{\text{oop}}$, $\Delta\alpha$ is obtained as a function of composition as shown in Fig. 3f, in which the composition dependence of the $|\lambda_{100}|$ value is also plotted. One can see that the composition dependences of $\Delta\alpha$ and $|\lambda_{100}|$ are fairly in agreement with each other. Moreover, a correlation is also suggested between $\alpha_{\text{oop}}$ and $|\lambda_{100}|$ except for the Mn or Co rich region. The values of $H_{\text{inh}}$ are shown in the Supplementary Information.

**Origins of correlation between magnetostriction and magnetic damping**

The finding of correlation between $|\lambda_{100}|$ and $\alpha$ in the magnetic nitrides, especially the strong correlation with extrinsic $\alpha$ allows us to deeply understand the underlying mechanisms of magnetostriction and magnetic damping. The origin of $\alpha$ is a long-standing issue[28] and the correlation between them, which was reported also for the other systems[29,30,31,32], has not been understood yet. In this study, we performed the first-principles calculations for $|\lambda_{100}|$ and intrinsic $\alpha$. Because the precise ferrimagnetic spin structure has not been determined for $Fe_{4-x}Mn_xN$, calculations were performed only for ferromagnetic $Fe_{4-y}Co_yN$. Fe and Co can exist at both of the two occupied sites[33,34] as shown in Fig. 1a.

Figure 4a shows the representative results for the density of states (DOS) for the majority and minority spins of $Fe_4N$, $Fe_{3.2}Co_{0.8}N$, and $Fe_2Co_2N$ around $E_F$. Note that the contribution from minority spin states much larger than majority spin especially around $E_F$. In addition, the peak shift toward lower energy side observed for minority spin states by replacing Fe with Co. The $y$ dependence of theoretical $M_s$ is given by the black solid circles in Fig. 4b and increasing Co content will lead to the reduction of total magnetic moment. However, the calculated $M_s$ is larger than the experimental values measured at 4 K (open star). Here, we notice that if the position of $E_F$ is shifted to the higher energy side, the calculated results are closer to the experimental results. This is simply explained due to the minority spin dominates the conduction band, hence shift toward higher energy side lead to decrease of the total magnetic moment. The red triangles and blue squares represent the results with $E_F$ shifted by 0.2 and 0.3 eV, respectively. The $E_F$ shift by 0.2 eV is appropriate because the $E_F$ shift by 0.3 eV gives rise to the calculated value lower than the experiment at the Co-rich composition. The $E_F$ position with the 0.2 eV shift is close to the peak top of the minority spin DOS of $Fe_4N$ as shown in Fig. 4a.

Using the DOS with the 0.2 eV $E_F$ shift and the Kamberský torque correlation model[35], intrinsic $\alpha$ values were calculated for $Fe_{4-y}Co_yN$. Figure 4c shows the calculated values of $\alpha$ ($\alpha_{cal}$) on the scattering rate ($\delta$), in which the intraband component is dominant in the small $\delta$ region whereas the interband component is dominant in the large $\delta$ region. From the resistivity of ~110 μΩ cm for $Fe_{4-y}Co_yN$[36], the magnitude of $\delta$ is estimated to be ~0.01 eV (denoted by the dotted line in Fig. 4c), suggesting that the intraband component is dominant for $Fe_4N$. Figure 4d compares $\alpha_{cal}$ at $\delta = 0.01$ eV with the experimental values of $\alpha_{oop}$ (intrinsic term of $\alpha$). The calculation qualitatively explains the trend of decreasing $\alpha_{oop}$ with increasing Co amount. The magnitude of $\alpha_{cal}$ is proportional to $D(E_F)^2$[37]. Thus, the $E_F$ shift away from the peak position of the minority spin DOS of $Fe_4N$ leads to the decrease in magnetic damping.

Figure 4e shows the total energy ($E_{tot}$) and magnetic anisotropy energy (MAE) of $Fe_4N$ as a function of the lattice constant perpendicular to the film surface ($c$) without modulation of the valence electron number ($N_v$). Assuming $E_{tot} = gc^2 + hc + i$, where $g$, $h$, and $i$ are constants, $\lambda_{100}$ can be calculated from the following equation[38]:

$$\lambda_{100} = -\frac{2}{3h} \cdot \frac{d(\text{MAE})}{dc}. \tag{1}$$

Figure 4f shows the $y$ dependence of calculated $\lambda_{100}$ for $Fe_{4-y}Co_yN$ using equation (1), where the black circles represent the results without modulating $N_v$ and the blue triangles represent the results in case that $N_v$ is increased by 0.4 / f.u. The experimental values are also given by the green squares for comparison. By increasing $N_v$ as in the case of calculation for $\alpha$, the calculated results are fairly in agreement with the experimental results. One sees that the experimentally obtained composition dependences of $\alpha$ and $\lambda_{100}$ for Fe$_{4-y}$Co$_y$N (Figs. 4d and 4f) are well explained by the first-principles calculations. It is important to note that the sign and magnitude of $\lambda_{100}$ strongly depend on the derivative of MAE on the lattice parameter $c$ (d(MAE)/d$c$). Using second order perturbation analysis (see Supplementary Fig. 3), we clarified that the dominant contribution to the d(MAE)/d$c$ is given by MAE from the spin conserving term in the minority-spin state, implying the crucial role of minority $D(E_F)$. Therefore, $D(E_F)$ plays a pivotal role to correlate between intrinsic $\alpha$ and $|\lambda_{100}|$.

In addition to the discussion about the correlation between "intrinsic" $\alpha$ and $|\lambda_{100}|$, we can discuss the correlation between "extrinsic" $\alpha$ and $|\lambda_{100}|$ observed experimentally. The extrinsic $\alpha$ is partly attributable to two-magnon scattering. A previous paper[22] mentioned that magnetostriction coupled with nonuniform stresses is a source of effective field for additional spin wave dynamics, which enhances two-magnon scattering and increases the extrinsic $\alpha$. Figure 5a displays the cross-sectional high-angle annular dark-field scanning transmission electron microscope (HAADF-STEM) image of the STO(001)/Fe$_4$N. The inset is a fast Fourier transform (FFT) image. The image shown in Fig. 5b is obtained by selecting appropriate diffraction spots from the FFT image and applying inverse FFT. The yellow marks denote the misfit dislocations, and the inhomogeneous strain is introduced around them. Since the lattice mismatches between STO and $Fe_{4-y}Co_yN$ do not change significantly (−2.8 and −4.0% for $y =$ 0 and 4, respectively), we consider the misfit dislocation density hardly depends on $y$. Therefore, the extrinsic $\alpha$ is enhanced in a material with large $|\lambda_{100}|$, and is also a useful indicator to investigate the magnitude of magnetostriction constant.

**Giant tunability and bipolarity of magnetoelasticity**

We demonstrated that the large negative $\lambda_{100}$ can be achieved for the epitaxially grown $Fe_4N$ film and the magnetostriction properties can be significantly varied by partially replacing Fe with Co or Mn. The magnitude of $\lambda_{100}$ was modulated in a wide range from −121 to +46 ppm. This fact means the giant tunability and bipolarity of magnetostriction in the ferromagnetic nitride. We found that the correlation between $\alpha$ and $|\lambda_{100}|$ experimentally and theoretically. The first-principles calculations indicated that $D(E_F)$ was the leading term to determine the magnitudes for both intrinsic $\alpha$ and $\lambda_{100}$, suggesting that appropriate control of $D(E_F)$ is indispensable for developing giant magnetoelastic materials. As for the correlation between extrinsic $\alpha$ and $|\lambda_{100}|$, the magnetostriction is a source of the two-magnon scattering, resulting in the large extrinsic $\alpha$ with increasing $\lambda_{100}$. We believe that the findings obtained on the $Fe_4N$ research platform can give a guiding principle for exploring new giant magnetoelastic materials.

We address several advantages of ferromagnetic nitrides as a material for spintronic strain sensors. The giant tunability and bipolarity of magnetostriction are useful to improve the sensitivity of strain detection. The Fe$_4$N with large negative $\lambda_{100}$ exhibits the alignment of magnetic moments along the orthogonal direction to the strain direction whereas the Fe$_{2.3}$Co$_{1.7}$N with large positive $\lambda_{100}$ exhibits the moment alignment along the strain direction. For GMR/TMR-based strain sensors with two or more ferromagnetic layers, these characteristics are essential to obtain the large variation in device resistance determined by the relative angles between the adjacent ferromagnetic layers under the strain application. We expect that the large difference in magnetostriction between the Fe$_4$N and the Fe$_{2.3}$Co$_{1.7}$N significantly improves the device performance. As mentioned in the introductory part, the Fe$_4$N shows the TMR by combining with a standard tunnel barrier material MgO, and the epitaxial Fe$_4$N layer can be grown on the flexible substrate. Also, we may realize the large-scaled device integration thanks to the fact that Fe$_4$N does not contain minor metal elements and rare earth elements. Therefore, we conclude that magnetic nitrides are leading candidate materials for high-sensitive spintronic strain sensors.

**Methods**

**Sample fabrication**

The nitride films with a thickness of approximately 23 nm were grown on STO(001) substrates using molecular beam epitaxy (MBE). STO substrates were ultrasonically cleaned in the order of acetone, ethanol, and deionized water, and then dipped in buffered hydrofluoric acid. Fe and Co were supplied by using electron beam guns, Mn by a Knudsen Cell, and N by a radio-frequency (rf) radical source (Eiko engineering: ER-1000). The deposition rate was fixed at 0.2 Å/s for the total of Fe+Co and Fe+Mn, and the composition ratio was modulated by changing the supply rate ratio. High purity (6N) $N_2$ gas was supplied at 1.0 sccm and the rf input power was fixed at 300 W. The deposition temperature was set at 450 °C. For the $Fe_{4-y}Co_yN$ ($y = 1.7$ and 2.2) films, the temperature was lowered to 400 °C to prevent nitrogen desorption. Subsequently, a 2-nm-thick Ti capping layer was formed on the nitride layers in an MBE chamber to prevent surface oxidation.

**Structural characterization**

The structures of the samples were characterized by XRD with a Cu-Kα radiation source (SmartLab: Rigaku Corp.). The film thicknesses were evaluated by an x-ray reflectivity method. The surface morphologies and crystal growth orientations were also monitored by reflection high-energy electron diffraction during the film deposition. The microstructure of the STO/Fe$_4$N was observed by HAADF-STEM (JEM-ARM200F: JEOL Ltd.). The composition ratios of the nitride films were evaluated using a field emission electron probe microanalyzer (JXA-8530F: JEOL Ltd.) with the standard samples for calibration.

**Magnetic property measurement**

The magnetization curves at room temperature were measured using a vibrating sample magnetometer (TM-VSM2614HGC-KIT, Tamakawa Co., Ltd.). The values of $M_s$ at 4 K were measured employing a superconducting quantum interference device magnetometer (MPMS3: Quantum Design Inc.). The magnetostriction was evaluated using an optical cantilever system[30] (Toei Scientific Industrial Co., Ltd.). FMR spectra were measured at room temperature using a vector network analyzer (N5222A: Agilent Technology) and coplanar wave guides. Frequency-domain FMR measurements were carried out, sweeping rf current in the range of 1 ~ 20 GHz and varying $H$. For the IP configuration measurements, an rf-compatible prober station (TKSPHE-3V10: Toei Scientific Industrial Co., Ltd.) was used with $200 \leq H \leq 2400$ Oe applied in the in-plane [100] direction of the films. For the OOP configuration measurements, a physical property measurement system (PPMS: Quantum Design Inc.) was used and $13.0 \leq H \leq 22.5$ kOe was applied in the perpendicular direction to the film surface.

**First-principles calculation**

We performed first-principles density functional calculation using VASP[39] to obtain electronic structures and MAE of $Fe_{4-y}Co_yN$ series. The projector augmented-wave (PAW) potential was used to describe the behavior of core electrons[40]. The generalized gradient approximation proposed by Perdew, Burke, and Ernzerhof was adopted for the exchange and correlation energies[41]. A unit cell containing 4 Fe (Co) and 1 N atoms was constructed. The virtual crystal approximation (VCA) was used to describe the mixing between Fe and Co atoms[42]. We used the plane-wave cutoff energy of 350 eV for the wave function expansion and 30×30×30 k-point mesh for wave vector integration in the first Brillouin zone. The damping was calculated using Torque Correlation Model by Kamberský[35]. The MAE was calculated by including spin-orbit interaction self-consistently. The magnetostriction constant was estimated by calculating MAE for a particular range of the lattice constant $c$ with fixing the volume of unit cell.

**Data availability**

The data that support the findings of this study are available from the corresponding authors upon reasonable request. Source data are provided with this paper.

**Code availability**

Computer codes used in this study are available from the corresponding author upon reasonable request.

**Acknowledgements**

The authors thank S. Kanai, S. Fukami, R. Y. Umetsu, and Y. Sakuraba for support to measure FMR, S. Mizukami and H. Kurebayashi for their valuable comments, I. Narita for technical support to do the composition analysis, and K. Omura and Y. Murakami for help to perform the XRD measurements. This work was partly supported by the Collaborative Research Center on Energy Materials, Institute for Materials Research (IMR), Tohoku University, and the Cooperative Research Project of the Research Institute of Electric Communication, Tohoku University. The magnetization curve and XRD measurements were carried out at the Cooperative Research and Development Center for Advanced Materials, IMR, Tohoku University. The first-principles calculations of the magnetic damping and the magnetostriction were performed on a Numerical Materials Simulator at NIMS. This work was supported by JSPS KAKENHI Grant Numbers JP22K18894 and JP22H04966, and Initiative to Establish Next-generation Novel Integrated Circuits Centers (X-nics) under Grant JPJ011438, the Ministry of Education, Culture, Sports, Science and Technology (MEXT).

**Author contributions**

Film growth, XRD measurements, and magnetization curve measurements were carried out by K.I. K.I. and T.S. performed magnetostriction measurements with support from Y.E. K.I. and T.S. performed FMR measurements. Y.S. carried out TEM observations. First-principles calculations were carried out by I.K. with support from Y.M. K.I. and T.S. wrote the draft. All the authors discussed the results and commented on the manuscript. T.S. conceived the project.

**Competing interests**

The authors declare no competing interests.

**Correspondence and requests for materials** should be addressed to Keita Ito.

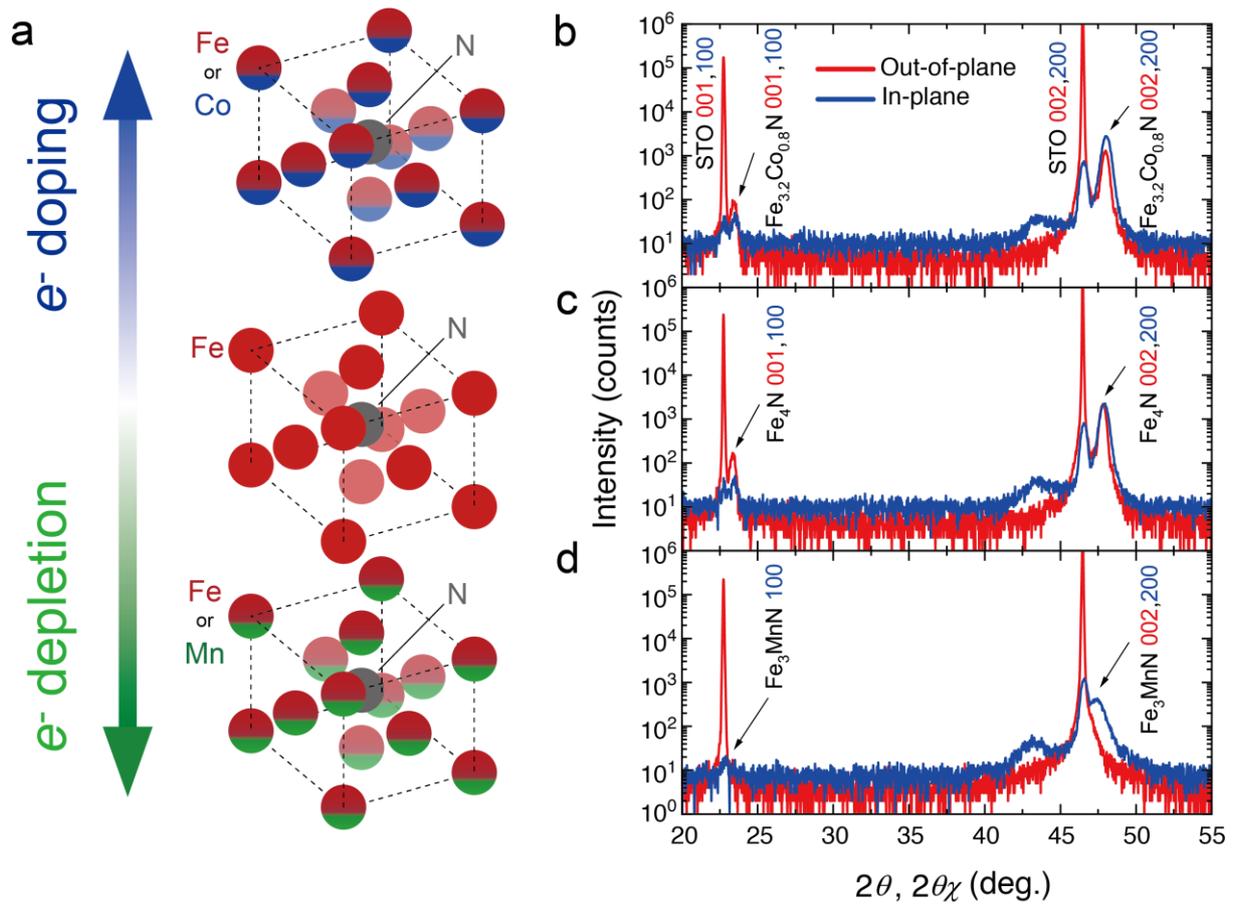

**Fig. 1 | Concept of element substitutions and structural analysis for magnetic nitrides.** (a) Crystal structures of magnetic nitrides; Co substitution into $Fe_4N$ corresponds to electron doping and Mn substitution to hole doping. Out-of-plane and in-plane XRD patterns of the (b) $Fe_{3.2}Co_{0.8}N$, (c) $Fe_4N$, and (d) $Fe_3MnN$ films. Red and blue lines correspond to the results with the out-of-plane and in-plane configurations, respectively.

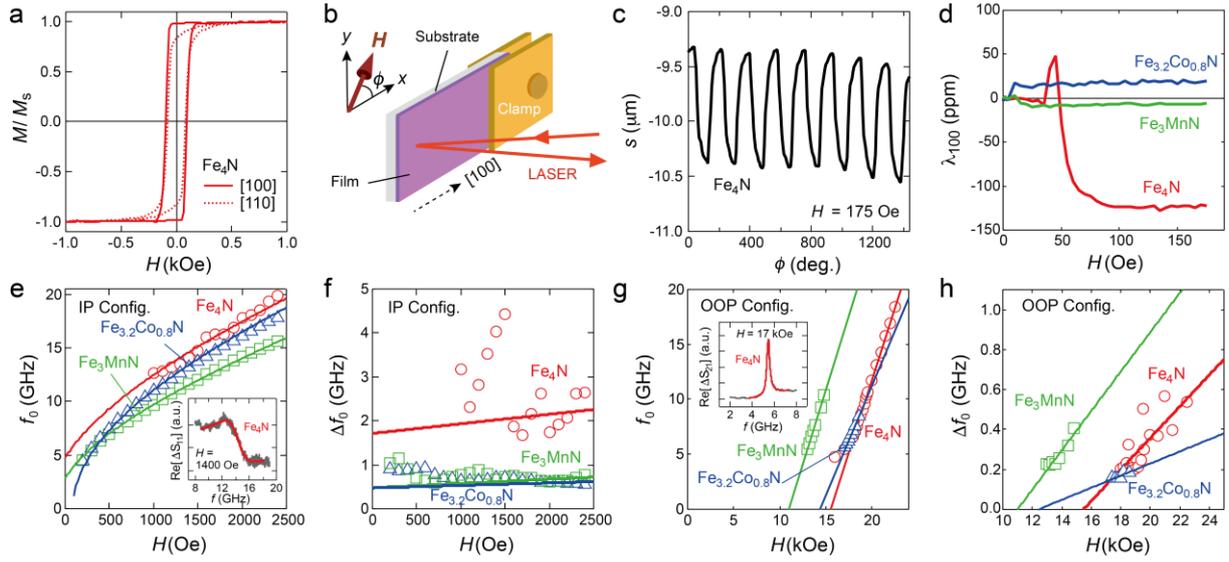

**Fig. 2 | Magnetization, magnetostriction and magnetization dynamics of magnetic nitrides.**

(a) In-plane magnetization curves of the $Fe_4N$ film measured at room temperature, where $M$ is normalized by $M_s$. The solid (dashed) curve denotes the result with $H$ applied in the in-plane [100] direction (in-plane [110] direction). (b) Illustration of experimental setup of an optical cantilever method. (c) $\phi$-dependence of $s$ at a rotational $H$ of 175 Oe applied to the $Fe_4N$ film. (d) $H$ dependence of $\lambda_{100}$ for the $Fe_4N$ (red line), $Fe_{3.2}Co_{0.8}N$ (blue line), and $Fe_3MnN$ (green line) films. (e) $H$ dependence of $f_0$ and (f) $\Delta f_0$ for the $Fe_4N$ (red line), $Fe_{3.2}Co_{0.8}N$ (blue line), and $Fe_3MnN$ (green line) films with the IP configuration. The inset of (e) shows the resonance spectrum together with the fitting result for the $Fe_4N$ film measured at $H = 1400$ Oe. (g) $H$ dependence of $f_0$ and (h) $\Delta f_0$ for the $Fe_4N$ (red line), $Fe_{3.2}Co_{0.8}N$ (blue line), and $Fe_3MnN$ (green line) films with the OOP configuration. The inset of (g) shows the resonance spectrum together with the fitting result for the $Fe_4N$ film measured at $H = 17$ kOe.

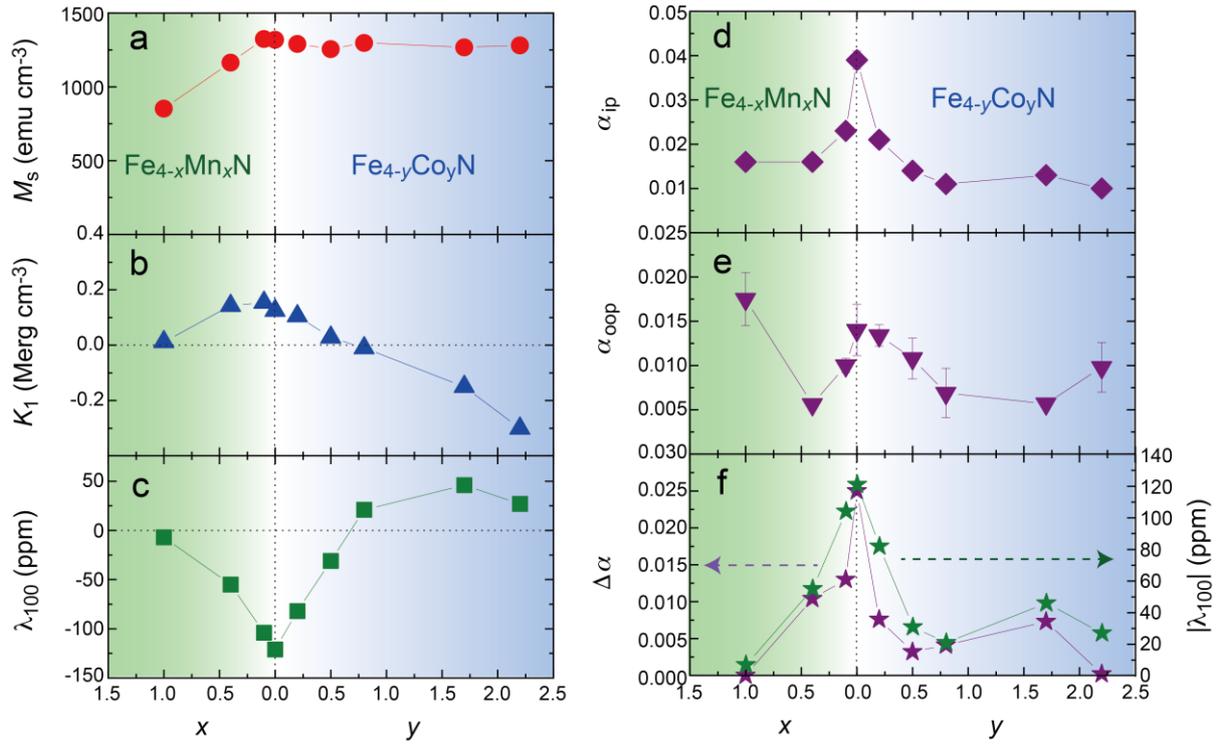

**Fig. 3 | Composition dependences of magnetic properties.** Composition dependences of (a) $M_s$, (b) $K_1$, (c) $\lambda_{100}$, (d) $\alpha_{ip}$, (e) $\alpha_{oop}$, and (f) $\Delta\alpha$ together with $|\lambda_{100}|$ for the $Fe_{4-x}Mn_xN$ and $Fe_{4-y}Co_yN$ films.

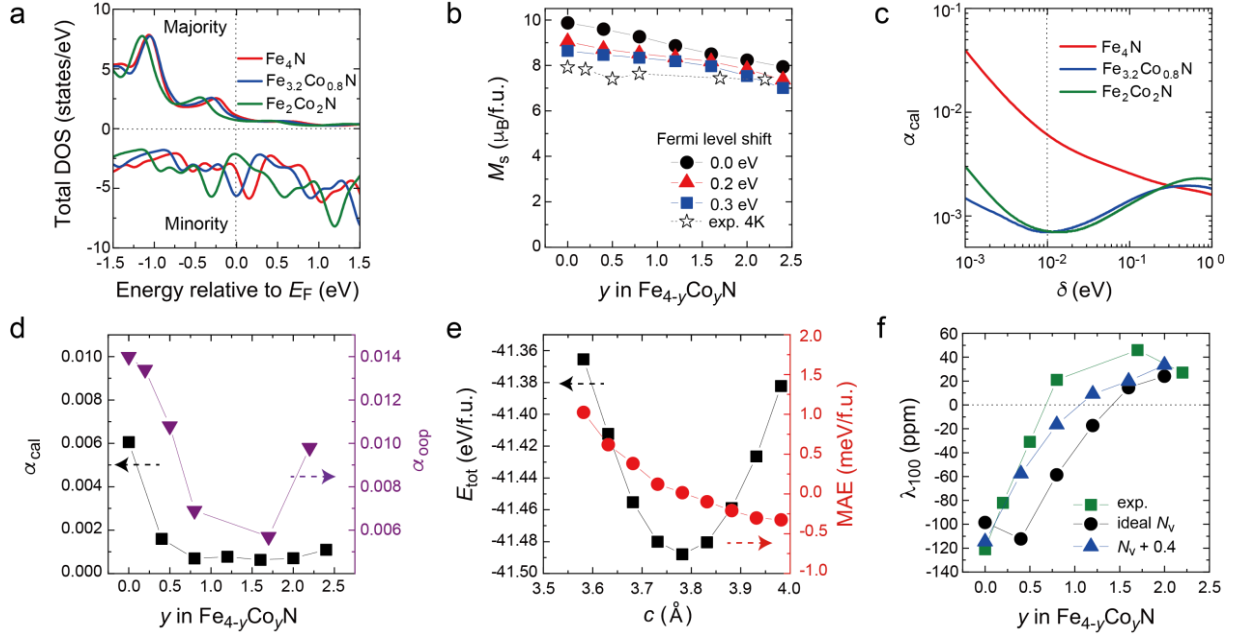

**Fig. 4 | First-principles calculations on electronic structures and magnetic properties of Fe$_{4-y}$Co$_y$N.** (a) Density of states (DOS) for the majority and minority spins of Fe$_4$N (red line), Fe$_{3.2}$Co$_{0.8}$N (blue line), and Fe$_2$Co$_2$N (green line) around $E_F$ predicted by first-principles calculations. (b) Composition dependence of theoretical and experimental $M_s$ values in Fe$_{4-y}$Co$_y$N. The black circles, red triangles, and blue squares represent the calculation results of shifting $E_F$ by 0, 0.2, and 0.3 eV, respectively. The open stars indicate the experimental $M_s$ values measured at 4 K. (c) Relationship between $\alpha_{cal}$ and $\delta$ for Fe$_4$N, Fe$_{3.2}$Co$_{0.8}$N, and Fe$_2$Co$_2$N. The position of $\delta = 0.01$ eV is denoted by the dotted line. (d) Composition dependence of $\alpha_{cal}$ at $\delta = 0.01$ eV (black squares) and $\alpha_{oop}$ (purple inverted triangles). (e) Dependence of $E_{tot}$ (black squares) and MAE (red circles) of Fe$_4$N on $c$ without modulation of $N_v$. (f) Calculated $y$ dependence of $\lambda_{100}$ for Fe$_{4-y}$Co$_y$N. The black circles and blue triangles show the results when $N_v = 0$ (no $E_F$ shift) and 0.4 / f.u., respectively. The green squares represent the

experimental results.

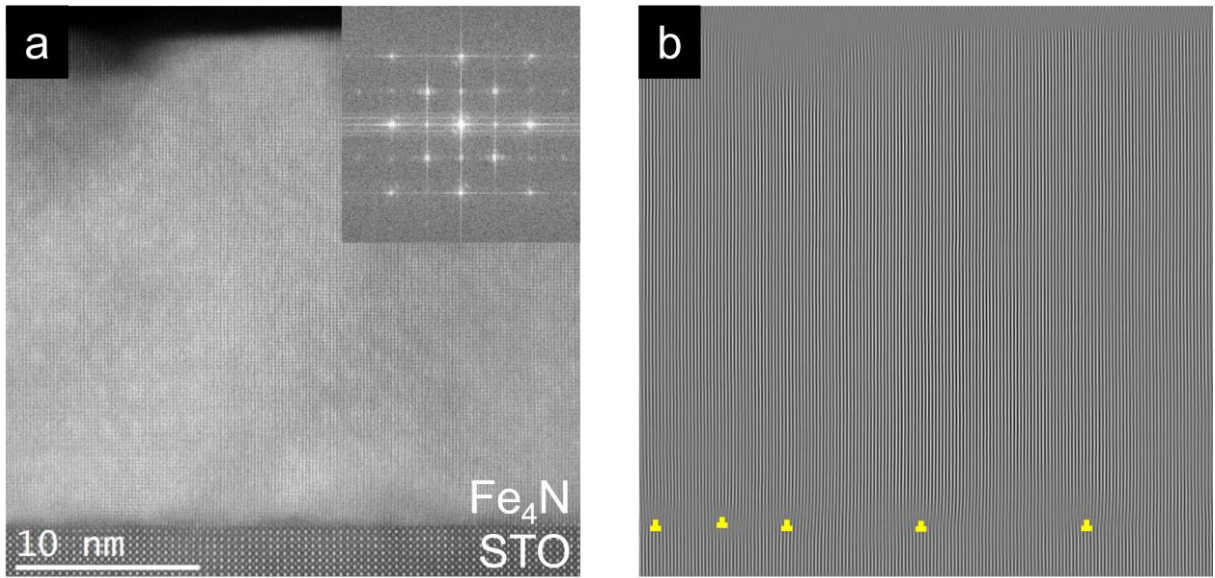

**Fig. 5 | Microstructure of the Fe$_4$N film on the STO substrate.** (a) Cross-sectional HAADF-STEM image of the STO(001)/Fe$_4$N. The inset shows a FFT image. (b) Inverse FFT image obtained from the inset of (a). The yellow marks show the misfit dislocations.

## Supplementary Information

Supplementary Fig. 1 shows the x-ray reflectivity (XRR) profile for the $Fe_4N$ film. The black and red lines correspond to the obtained data and the fitting line, respectively. Supplementary Table 1 summarizes the film thicknesses of the nitride films estimated from the XRR. The accuracy of the estimated thicknesses is high because the fitting was done with high precision.

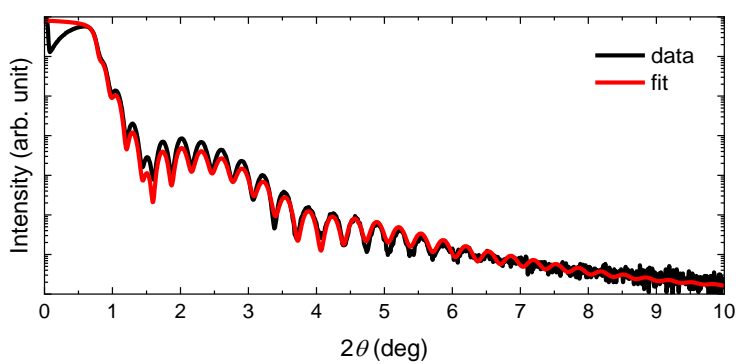

**Supplementary Fig. 1**   XRR profile of the $Fe_4N$ film.

**Supplementary Table 1** The thicknesses of the nitride films estimated from the XRR.

| Samples | Thickness (nm) |
|:---:|:---:|
| $Fe_4N$ | 27 |
| $Fe_{3.9}Mn_{0.1}N$ | 21 |
| $Fe_{3.6}Mn_{0.4}N$ | 25 |
| $Fe_3MnN$ | 25 |
| $Fe_{3.8}Co_{0.2}N$ | 22 |
| $Fe_{3.5}Co_{0.5}N$ | 21 |
| $Fe_{3.2}Co_{0.8}N$ | 25 |
| $Fe_{2.3}Co_{1.7}N$ | 21 |
| $Fe_{1.8}Co_{2.2}N$ | 22 |

According to the model for the optical cantilever method, the displacement of reflected position of the laser light ($\Delta s$) is expressed by the following equation[20]:

$$\Delta s = C\lambda_{100} \cdot \frac{h_{\text{film}}}{h_{\text{sub}}^2} \cdot \frac{E_{\text{film}}}{E_{\text{sub}}} \cdot \frac{(1+\nu_{\text{sub}})}{(1+\nu_{\text{film}})}, \qquad \text{(Supplementary Eq. 1)}$$

where $C$ is the coefficient determined by the measurement system geometry, $\lambda_{100}$ is the magnetostriction constant along [100] direction, $h_{\text{sub}}$ ($h_{\text{film}}$) is the substrate (film) thickness, $E_{\text{sub}}$ ($E_{\text{film}}$) is the Young's modulus of the substrate (film), and $\nu_{\text{sub}}$ ($\nu_{\text{film}}$) is the Poisson's ratio of the substrate (film). Supplementary Table 2 summarizes $E$ and $\nu$ values of $SrTiO_3$[S1], $Fe_4N$[S2], $Fe_3MnN$[S3], $Fe_3CoN$[S4], and $Co_4N$[S5]. For the evaluation of $\lambda_{100}$ for all the samples, the $E$ and $\nu$ values of $Fe_4N$ were used because accurate $E$ and $\nu$ values for each composition are unknown. The $\nu$ values of $Fe_4N$, $Fe_3MnN$, $Fe_3CoN$, and $Co_4N$ are close, while the $E$ values of $Fe_3MnN$, $Fe_3CoN$, and $Co_4N$ are larger than that of $Fe_4N$. This means that obtained $\lambda_{100}$ values of the $Fe_{4-x}Mn_xN$ and $Fe_{4-y}Co_yN$ films are slightly overestimated. However, the overestimation of $\lambda_{100}$ does not affect the main conclusion of this study.

**Supplementary Table 2** $E$ and $\nu$ values of SrTiO$_3$, Fe$_4$N, Fe$_3$MnN, Fe$_3$CoN, and Co$_4$N.

| Materials | $E$ (GPa) | $\nu$ |
|---|---|---|
| SrTiO$_3$ | 278 | 0.24 |
| Fe$_4$N | 162 | 0.36 |
| Fe$_3$MnN (Ferromagnetic) | 202 | 0.31 |
| Fe$_3$CoN | 217 | 0.32 |
| Co$_4$N | 239 | 0.31 |

As mentioned in the main text, the fitting functions for the resonance linewidth ($\Delta f_0$) involve the inhomogeneous magnetic field ($H_{inh}$): $\Delta f_0(H) = \alpha_{ip}(\gamma/2\pi)(2H + H_{a,ip} + 4\pi M_{eff}) + (\gamma/2\pi)H_{inh}$ for the IP configuration, and $\Delta f_0(H) = \alpha_{oop}(\gamma/2\pi)(2H - 2H_s) + (\gamma/2\pi)H_{inh}$ for the OOP configuration. Thus, we evaluated the values of $H_{inh}$ together with $\alpha_{ip}$ and $\alpha_{oop}$. In the case of IP configuration, the measured values of $\Delta f_0$ are remarkably scattered as seen in Fig. 2f, which gives rise to the estimated values of $H_{inh}$ with significant errors, *e.g.* $H_{inh} = 113 \pm 506$ Oe for $x = 0.5$. Thus, we cannot discuss the values of $H_{inh}$ for the IP configuration because of low accuracy of estimation. In contrast to the IP configuration, the OOP configuration leads to the values of $H_{inh}$ with relatively small errors. Supplementary Fig. 2 plots the composition dependence of $H_{inh}$ estimated in the OOP configuration. One sees that the values of $H_{inh}$ are negligibly small and do not show a clear composition dependence.

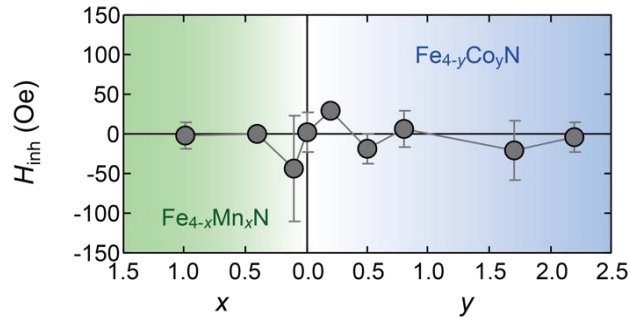

**Supplementary Fig. 2**  Composition dependence of inhomogeneous magnetic field ($H_{inh}$) estimated in the out-of-plane configuration.

Supplementary Fig. 3 shows the spin-resolved magnetocrystalline anisotropy energies (MAE) for $Fe_4N$ and $Co_4N$ in the second-order perturbation of spin-orbit interaction as a function of $c$ lattice parameter. We clarified that the dominant contribution to the d(MAE)/d$c$ is given by competing term of negative trend of dMAE($\downarrow\Rightarrow\downarrow$)/d$c$ and positive trend of d(MAE($\uparrow\Rightarrow\downarrow$)/d$c$, where $\sigma \Rightarrow \sigma'$ indicates the spin-transition process in the second order perturbation from the initial spin $\sigma$ state to the intermediate spin $\sigma'$ state. For $Fe_4N$, the former is much more dominant than the latter, hence give the large negative magnetostriction. This means that the anisotropy of orbital moment of Fe related to the MAE($\downarrow\Rightarrow\downarrow$) significantly contributes to the magnetostriction of $Fe_4N$. However, the latter is slightly more dominant for $Co_4N$. Nevertheless, both cases implying the crucial role of minority-spin $D(E_F)$. Therefore, $D(E_F)$ plays a major role to correlate between intrinsic $\alpha$ and $|\lambda_{100}|$.

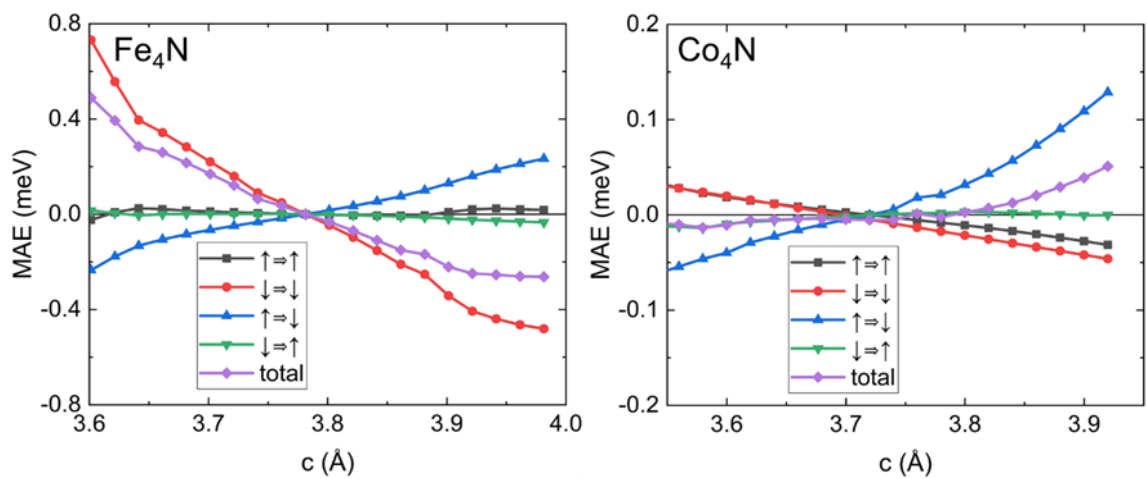

**Supplementary Fig. 3** Spin-resolved MAE for Fe$_4$N and Co$_4$N in the second-order perturbation of spin-orbit interaction as a function of $c$ lattice parameter.